# Spin-orbit torques in Co/Pd multilayer nanowires

Mahdi Jamali,[1] Kulothungasagaran Narayanapillai,[1] Xuepeng Qiu,[1] Li Ming Loong,[1] Aurelien Manchon,[2] and Hyunsoo Yang[1,*]

[1]Department of Electrical and Computer Engineering and NUSNNI, National University of Singapore, 117576, Singapore

[2]Division of Physical Science and Engineering, King Abdullah University of Science and Technology (KAUST), Thuwal 23955, Saudi Arabia

Current induced spin-orbit torques have been studied in ferromagnetic nanowires made of 20 nm thick Co/Pd multilayers with perpendicular magnetic anisotropy. Using Hall voltage and lock-in measurements, it is found that upon injection of an electric current both in-plane (Slonczewski-like) and perpendicular (field-like) torques build up in the nanowire. The torque efficiencies are found to be as large as 1.17 kOe and 5 kOe at $10^8$ A/cm$^2$ for the in-plane and perpendicular components, respectively, which is surprisingly comparable to previous studies in ultrathin (~ 1 nm) magnetic bilayers. We show that this result cannot be explained solely by spin Hall effect induced torque at the outer interfaces, indicating a probable contribution of the bulk of the Co/Pd multilayer.

The manipulation of the magnetization in ferromagnetic metallic systems through current induced spin transfer torque in metallic spin-valves and tunneling systems has been studied for the past fifteen years [1-4]. Recently, the study of current induced spin-orbit torques in ultrathin magnetic structures in the absence of a spin polarizer has attracted a strong interest among researchers due to its potential for low power magnetization switching [5-22]. Different phenomena, such as the Rashba effect [5, 23, 24] and the spin Hall effect [6], have been proposed to explain the current induced torques with an in-plane current. However, these theories lack accurate quantitative predictions and no unified scheme is currently available. It is also possible that bulk effects such as the spin Hall or spin swapping effect [25] and interfacial effects (such as, but not limited to, the Rashba torques) coexist. Therefore, a reliable characterization technique that can quantify the relative intensity of the torques induced by the electric current is important to understand the underlying mechanisms.

In the recent reports of the current induced torques, the thickness of the ferromagnetic material was ~ 1 nm, which is impractical for real applications [7, 13]. In addition, at such thin thicknesses the complex microscopic structure renders the theoretical modeling of the experimental conditions rather challenging [26]. Furthermore, it has been found that by increasing the thickness of the ferromagnet above 1.5 nm, the intensity of the current induced spin-orbit field decreases by an order of magnitude [10, 14]. In this letter, we show that the current induced spin-orbit torques from Co/Pd multilayer nanowires can be comparable to the current induced torque in the ultrathin layer counterparts. The experimental data were obtained using a homodyne detection of the magnetization dynamics driven by a combination of ac and dc currents. Based on macrospin modeling, it is found that the relative intensity of the transverse (field-like) to the longitudinal (spin transfer-like) torque is 4.3 in our Co/Pd structure, whereas the magnitude of torque is comparable to the largest value obtained to date in ultrathin bilayers. Using the drift-diffusion approach, we show that the spin Hall effect alone is not sufficient to explain our results and that important contributions from the inner Co/Pd interfaces should probably be taken into account. This observation opens promising directions towards the development of spin-orbit torque in bulk systems.

For the study of the current induced torques in the nanowire, we have utilized a lock-in technique [10, 13] with dc and ac currents to characterize the devices at different bias points. The nanowire has a perpendicular magnetic anisotropy subjected to a small perturbation induced either by a current or by an in-plane magnetic field. In order to derive an expression for the different torque terms, we consider a macrospin approach with a perpendicular anisotropy along the $z$-axis, leading to a torque $\hat{\tau} = \gamma\tau_{\|}\hat{m} \times (\hat{y} \times \hat{m}) + \gamma\tau_{\perp}\hat{m} \times \hat{y}$, where $\hat{y}$ is the direction transverse to the injected current and in the film plane of the magnetic structure, and $\gamma$ is the gyromagnetic ratio. The components $\tau_{\|}$ and $\tau_{\perp}$ are the in-plane and perpendicular torque terms, respectively. By assuming linear response, the torques can be written as $\tau_{\|,\perp} = \beta_{\|,\perp}\left(I_{dc} + I_{ac}\sin\omega t\right)$, where $I_{dc}$ and $I_{ac}$ are the amplitudes of the dc and ac components of the injected current, respectively, and $\beta_{\|}$ and $\beta_{\perp}$ are the intensity of the longitudinal and transverse torques. The modified Landau-Lifshitz-Gilbert equation in the presence of the current induced torques can be written as

$$\frac{d\hat{m}}{dt} = -\gamma\hat{m} \times \left(H_{\perp}m_z\hat{z} + H_z\hat{z} + H_x\hat{x}\right) + \alpha\hat{m} \times \frac{d\hat{m}}{dt} + \gamma\tau_{\|}\hat{m} \times \left(\hat{y} \times \hat{m}\right) - \gamma\left(H_y - \tau_{\perp}\right)\hat{m} \times \hat{y} \quad (1)$$

where $H_x$, $H_y$, and $H_z$ are the components of the bias magnetic field and $H_\perp = H_K - 4\pi M_s$, where $H_K$ is the anisotropy field and $M_s$ is the saturation magnetization. The Hall voltage is $V_H = \left[\rho_{AHE}\cos\theta + \rho_{PHE}\sin^2\theta\sin\phi\cos\phi\right]I_\omega$, where $(\theta,\phi)$ are the polar and azimuthal angles of the magnetization direction. The first term refers to the anomalous Hall effect (AHE) and the second term is the planar Hall effect (PHE) [27]. Since the lock-in frequency (< 1 kHz) is much smaller than the typical frequency of the magnetization dynamics (~ GHz), the torques are considered to be constant on the timescale of the magnetization dynamics. Furthermore, we assume that the spin torques and external magnetic field are small compared to the magnetic anisotropy, so that $\theta \approx 0, \pi$. Writing Eq. (1) in the spherical coordinate system and eliminating time dependent magnetization terms, the Hall signal adopts the form $V_H = \varepsilon\left[C_0 + C_\omega \sin\omega t + C_{2\omega}\cos 2\omega t\right]$, where

$$C_0 = \left(\rho_{AHE}I_{dc} + \left[2\rho_{PHE}\beta_\perp\beta_\parallel - \rho_{AHE}\left(\beta_\parallel^2 + \beta_\perp^2\right)\right]\left(I_{dc}^2 + 3\frac{I_{ac}^2}{2}\right)\frac{I_{dc}}{2H_\perp^2}\right)$$
$$C_\omega = \left(\rho_{AHE}I_{ac} + 3\left[2\rho_{PHE}\beta_\perp\beta_\parallel - \rho_{AHE}\left(\beta_\parallel^2 + \beta_\perp^2\right)\right]\left(I_{dc}^2 + \frac{I_{ac}^2}{4}\right)\frac{I_{ac}}{2H_\perp^2}\right) \quad (2)$$
$$C_{2\omega} = -3\left[2\rho_{PHE}\beta_\perp\beta_\parallel - \rho_{AHE}\left(\beta_\parallel^2 + \beta_\perp^2\right)\right]\frac{I_{dc}I_{ac}^2}{4H_\perp^2}$$

are the dc, the first harmonic, and the second harmonic components, respectively. $\varepsilon = \pm 1$ corresponds to $\theta = 0, \pi$. Note that Eq. (2) is derived in the absence of external magnetic field. General expressions can be found in Supplementary Materials [28]. In order to determine $\beta_\parallel$ and $\beta_\perp$, two sets of measurements have been performed. In the first case, no magnetic field has been applied to the sample during the measurement ($H_x = H_y = H_z = 0$). In this case, the relation $2\rho_{PHE}\beta_\perp\beta_\parallel - \rho_{AHE}\left(\beta_\parallel^2 + \beta_\perp^2\right)$ can be extracted from Eq. (2). In the second case, a small magnetic field has been applied along the *x*-axis and the relation $\rho_{PHE}\beta_{\perp} - \rho_{AHE}\beta_{\parallel}$ has been extracted from the measurement results. Combining both measurements provides an estimation of the individual torque components. As seen in Eq. (2) the dc component of the Hall voltage is a function of the ac-current, which is related to the mixing of the ac signals [28].

Devices are fabricated by the sputter deposition of Ta (4 nm)/Ru (20 nm)/[Pd (0.7 nm)/Co (0.2 nm)]$_{22}$/Ta (4 nm) on a glass substrate, followed by the patterning of the nanowire using e-beam lithography and argon ion milling. A second photolithography process is used to pattern Ta (5 nm)/Cu (100 nm) contacts. Before contact deposition, the interface between the contacts and the ferromagnetic structure is etched through 2 nm to clean the interface. The transmission electron microscopy (TEM) image of the deposited film is shown in Fig. 1(a). The Co/Pd multilayer is clearly visible in the image. The scanning electron micrograph (SEM) image of the device structure is shown in Fig. 1(b). The width of the nanowire and the Hall bar varies between 500 to 2000 nm, and the length of the nanowire changes from 10 to 30 μm. The input current is injected between $A_1A_2$ electrodes, and the Hall signal is measured between $B_1B_2$ or $C_1C_2$ electrodes using a lock-in amplifier [28]. From the hysteresis loops of the Co/Pd film measured by a vibrating sample magnetometer (VSM), the anisotropy field ($H_K$) is 11.2 kOe and the coercivity of the thin film is ~ 1 kOe in the *z*-direction, as shown in Fig. 1(c). The output signal for a sinusoidal current with an amplitude of $I_{ac}$ = 1 mA is shown in Fig. 1(d). Both the first and second harmonics have the same switching field. $V_{FH}$ and $V_{SH}$ are the amplitude of the first and second harmonics at zero field, which are 3.3 μV and 21.2 nV, respectively.

In order to evaluate the torques, we have changed the input current, and measured $V_{FH}$ and $V_{SH}$ without any magnetic field. In Fig. 2(a), $V_{FH}$ is shown as a function of the ac currents for three different values of dc current. By increasing the ac current, $V_{FH}$, which corresponds to the anomalous Hall resistance, increases linearly. Furthermore, by applying $I_{dc} = \pm 1$ mA, no significant change in the amplitude of the first harmonic loop has been observed, as the first linear term dominates. Using Eq. (2), one can calculate the Hall resistivity ($\rho_{AHE}$ = 2.52 m$\Omega$). The planar Hall effect has been determined independently and estimated as $\rho_{PHE} = 17\rho_{AHE}$ [28]. We have also measured $V_{SH}$ as a function of the ac current as shown in Fig. 2(b). Upon application of $I_{dc}$ = 1 mA, corresponding to a current density of $\sim 4\times10^6$ A/cm$^2$, there is a drastic change in the second harmonic loop amplitude. From the curve fitting using Eq. (2), we have $2\rho_{PHE}\beta_\perp\beta_\parallel - \rho_{AHE}(\beta_\parallel^2 + \beta_\perp^2) \approx 6.68\times10^8 \Omega\cdot\text{Oe}^2/\text{A}^2$. We have also performed the first harmonic measurement at different dc currents at a fixed $I_{ac}$ = 1 mA as shown in Fig. 2(c), resulting in $2\rho_{PHE}\beta_\perp\beta_\parallel - \rho_{AHE}(\beta_\perp^2 + \beta_\parallel^2) = 7.3\times10^8 \Omega\cdot\text{Oe}^2/\text{A}^2$. $V_{SH}$ at different dc currents for $I_{ac}$ = 1 mA is shown in Fig. 2(d) and it provides a consistent result with that of Fig. 2(b).

For the characterization of the intensity of the individual torque terms, a bias magnetic field has been applied at $\theta = 7°$ away from the $z$-axis and in the $z$-$x$ plane along the nanowire. In this case $H_x$ becomes finite and the value of $\rho_{PHE}\beta_\perp - \rho_{AHE}\beta_\parallel$ can be obtained. In the subsequent measurements, the measured data are subtracted from the zero field measurement data. Figure 3(a) shows the changes in $V_{FH}$ as a function of the ac current at different bias magnetic fields for $I_{dc}$ = 1 mA. Similar to the results in Fig. 2(a), the first harmonic voltage increases linearly by increasing the amplitude of ac current. Furthermore, $V_{FH}$ increases as the applied magnetic field is increased. We have also measured $V_{FH}$ for different values of dc current while the ac current was fixed at 1 mA, as seen in Fig. 3(b). The first harmonic signal increases with increasing the bias field, and the behavior becomes nonlinear with respect to the polarity of the dc currents. The nonlinear behavior is attributed to the large angle deviation of the magnetization from the rest position.

Using the lowest order in excitation [28], $\partial C_\omega / \partial I_{ac}$ can be written as $\partial C_\omega / \partial I_{ac} \approx \rho_{AHE} + 2I_{dc}\left[\rho_{PHE}\beta_\perp - \rho_{AHE}\beta_\parallel\right] H\sin\theta / (H_\perp + H\cos\theta)^2$. In Fig. 3(c) the changes in the first harmonic signal at different bias fields, from the results in Fig. 3(a), have been plotted for $I_{dc}$, $I_{ac}$ = 1 mA with a curve fit. As seen in the inset of Fig. 3(c), the curve fitting deviates from experimental results due to domain wall nucleation for applied fields larger than 400-500 Oe, as can be seen in Fig. 1(d). When the bias field is less than 580 Oe, which corresponds to a transverse bias field of $H_x$ = 70.7 Oe, the first harmonic signal can be fitted well with the formula. From the fitting we find $\rho_{PHE}\beta_\perp - \rho_{AHE}\beta_\parallel \approx 9\times10^3$ Ω·Oe/A. Similarly $\partial C_\omega / \partial I_{dc}$ [28] can be written as $\partial C_\omega / \partial I_{dc} = 2I_{ac}\left[\rho_{PHE}\beta_\perp - \rho_{AHE}\beta_\parallel\right] H\sin\theta / (H_\perp + H\cos\theta)^2$. In Fig. 3(d), we have shown the changes in the first harmonic voltage at different bias fields, obtained from Fig. 3(b), for $I_{dc}$, $I_{ac}$ = 1 mA with a curve fit. As shown in the inset of Fig. 3(d), for the low bias magnetic fields ($H$ < 390 Oe), the fitted curve matches well with the experimental data, while at large fields ($H$ > 390 Oe) the fitted curve deviates from the measurement results, similar to Fig. 3(c). From the curve fitting we have $\rho_{PHE}\beta_\perp - \rho_{AHE}\beta_\parallel \approx 8\times10^3$ Ω·Oe/A, which is very close to the result in Fig. 3(c).

Solving the coupled equation for the longitudinal and transverse torque efficiencies gives $\beta_\| \approx 46.8(\pm 4)$ and $\beta_\perp \approx 201.2(\pm 11.7)$ Oe/mA. Note that the second unphysical solution providing $\beta_\| \approx -6.4 \times 10^3$ Oe/mA has been discarded. Hence, the ratio between the transverse and longitudinal torques is ~ 4.3: the transverse (field-like) torque is larger than the longitudinal (spin transfer-like) torque, even though there is no oxide layer in the structure. For comparison, we provide the torque efficiency from previous experiments in Table 1. In order to compare, we define a dimensionless electrical efficiency $\alpha_i = (2e/\hbar) M_s t \beta_i$ ($i = \|, \perp$) which corresponds to an effective spin Hall angle in the spin Hall effect interpretation. While the ratio $\beta_\perp / \beta_\|$ is similar to the previous reports [10, 27] (see Table 1), the dimensionless torque efficiency is, to the best of our knowledge, the largest reported to date.

Up until now, most of the experiments have been interpreted in terms of the spin Hall effect [26] or Rashba torque arising from the outer interfaces defining the structure inversion asymmetry of the ultrathin magnetic layer [11, 12, 26, 29]. In the present structure, the 20 nm thick Co/Pd multilayer is sandwiched between the top Ta and bottom Ru interfaces. Between these two interfaces, Ta is known to generate strong spin Hall effect [6], whereas Ru has a rather small spin-orbit contribution and is not expected to significantly contribute to the spin Hall effect. In order to identify the origin of the large spin-orbit torque measured above, we developed a drift-diffusion model of the spin Hall torque. The drift-diffusion model is in principle limited to systems larger than the electron mean free path and governed by spin accumulation (such as vertical spin-valves). It is, therefore, not adapted to model spin transports in ultrathin magnetic bilayers [16]. However, considering the Co/Pd multilayer as an effective homogeneous ferromagnet, the structure can be modeled as three layers, thicker than their mean free paths, where the spin transport induced by the spin Hall effect is perpendicular to the film plane. The dynamics is governed by the coupled equations $\hat{\nabla}^2 \hat{S} = \frac{1}{\lambda_J^2}\hat{S}\times\hat{m}+\frac{1}{\lambda_\varphi^2}\hat{m}\times(\hat{S}\times\hat{m})+\frac{1}{\lambda_{sf}^2}\hat{S}$ and $\hat{J}_i^s = -D(\hat{\nabla}S_i + \alpha_H \hat{e}_i \times \hat{\nabla}n)$, where the first equation describes the spin accumulation diffusion in terms of spin precession $\lambda_J$, dephasing $\lambda_\varphi$, and spin-flip diffusion length $\lambda_{sf}$. The second equation defines the spin current density in the presence of the spin Hall effect. $\hat{S}$ is the spin density, $n$ is the charge density, $D$ is the diffusion coefficient, and $\alpha_H$ is the spin Hall angle. In our model, we consider that the spin Hall effect is only present in the adjacent normal metal. Within the spin Hall model, the perpendicular torque arises from the non-adiabatic absorption of the injected spin in the presence of spin relaxation/dephasing [26]: the spin relaxation distorts the spin precession, which gives rise to a redistribution of the spin density, resulting in the emergence of a perpendicular torque component. Therefore, the perpendicular torque is controlled by the combined effect of spin precession versus spin relaxation/dephasing [28]. To obtain an analytical result, we assume the continuity of the transverse spin density and spin current at the interfaces, and short spin dephasing and precession lengths. This yields the fields corresponding to the in-plane and perpendicular torque efficiency [26]

$$\beta_\parallel^{SHE} = \frac{H_\parallel}{j_N} = \alpha_H \frac{\hbar}{2eM_s d_F}(C-1)\frac{(\chi_I^2+\chi_R^2)C+\chi_R S}{\chi_I^2 C^2 + [\chi_R C + S]^2}$$

$$\beta_\perp^{SHE} = \frac{H_\perp}{j_N} = -\alpha_H \frac{\hbar}{2eM_s d_F}(C-1)\frac{\chi_I S}{\chi_I^2 C^2 + [\chi_R C + S]^2}$$

where $C = \cosh d_N / \lambda_{sfN}, S = \sinh d_N / \lambda_{sfN}$, and $\chi_{R,I} = \lambda_{sfF} \sqrt{\frac{1}{2}\left(\sqrt{\left(\lambda_{sfF}^{-2} + \lambda_\varphi^{-2}\right)^2 + \lambda_J^{-4}} \pm \left(\lambda_{sfF}^{-2} + \lambda_\varphi^{-2}\right)\right)}$ .

Here, $d_{N,F}$ is the thickness of the normal (ferromagnetic) layer. Assuming a thick normal metal ($d_N \gg \lambda_{sfN}$) and reasonable parameters $\lambda_{sfF} = 10$ nm, $\lambda_J = \lambda_\varphi = 0.5$ nm, $M_s = 623$ emu/cc, and $d_F = 20$ nm, this provides a magnitude of the in-plane torque of $\beta_\parallel^{SHE} \approx \frac{\alpha_H \hbar}{2eM_s d_F} \cong \alpha_H \times 263$ Oe/$(10^8 \text{A/cm}^2)$, which is obviously much smaller than observed experimentally. Similarly, the field-like torque obtained through this model is only 1% of the in-plane torque. This analysis shows that the large torque observed in the Co/Pd multilayer cannot be explained by the spin Hall effect only and additional mechanisms should be brought to the physical picture.

The fact that the torques extracted from our measurements are in the *same order of magnitude* as that for systems one order of magnitude *thinner* (see Table 1) indicates that contributions from inner Co/Pd interfaces might take place. If one now adopts the phenomenology of Rashba torque $\beta_\perp^R = \alpha_R P / \mu_B M_s$, where the polarization is $P = 0.5$, this results in the Rashba splitting of $\alpha_R \approx 3.6 \cdot 10^{-11} eV \cdot m$, which is a reasonable magnitude [5]. Such an effective spin-orbit field can arise if two successive Co/Pd and Pd/Co interfaces are structurally dissimilar; otherwise their combined effect cancels out. The large lattice mismatch (~9%) between Pd and Co (respectively 3.89 Å and 3.55 Å in fcc structure [30]) results in strong lattice distortions at their interface. More importantly, this distortion is 30% stronger for Co/Pd than for Pd/Co interfaces [31], which is expected to impact the local band structure and explain the presence of a Rashba-like spin-orbit splitting in the bulk Co/Pd multilayer. Although the spin Hall effect alone cannot account for the large in-plane torque in our structure, the presence of dissymmetric lattice distortions may also result in the emergence of the recently proposed intrinsic spin-orbit torque [32]. Likewise, strong momentum scattering at the Pd/Co interfaces may significantly contribute to the in-plane torque [12, 29]. An accurate model of such a complex system is beyond the scope of the present study.

In summary, the current induced spin-orbit torques have been studied in Co/Pd multilayer nanowires. Using a lock-in technique, the first and second harmonics of the Hall voltage have been measured at different ac and dc currents. It is found that the longitudinal torque is $\sim 1.17 \times 10^{-8}$ kOe·cm$^2$/A and the ratio of the transverse to the longitudinal torque is $\sim 4.3$. The spin Hall effect is not sufficient to explain either the magnitude or the ratio of the torques. The experimental results indicate that thick magnetic multilayers can display large spin-orbit torques, which has important implications for future spintronic devices based on in-plane current induced switching.

We acknowledge fruitful discussions with I.M. Miron and K.J. Lee concerning the data analysis. M.J. and K.N. contributed equally to this work. This work was supported by the Singapore National Research Foundation under CRP Award No. NRF-CRP 4-2008-06.

* eleyang@nus.edu.sg

Figure Captions

Figure 1. (a) TEM image of a Co/Pd multilayer film. (b) SEM image of the device with the measurement scheme. (c) Hysteresis loop of the Co/Pd multilayer thin film before patterning. (d) The first and second harmonics of the Hall voltage measured for $I_{ac}$ = 1 mA.

Figure 2. (a) First harmonic loop amplitude versus ac-currents at different dc-currents. (b) Second harmonic amplitude as a function of the ac-current for $I_{dc}$ = 0, 1 mA with a curve fit. The first harmonic (c) and second harmonic (d) loop amplitude at different dc-currents for $I_{ac}$ = 1 mA with a corresponding curve fit.

Figure 3. The changes in the first harmonic amplitude as a function of ac (a) or dc (b) currents at different bias fields. The first derivative of the first harmonic loop amplitude with respect to the ac (c) or dc (d) currents as a function of bias fields with curve fits. The insets show results at a larger range of the bias field.

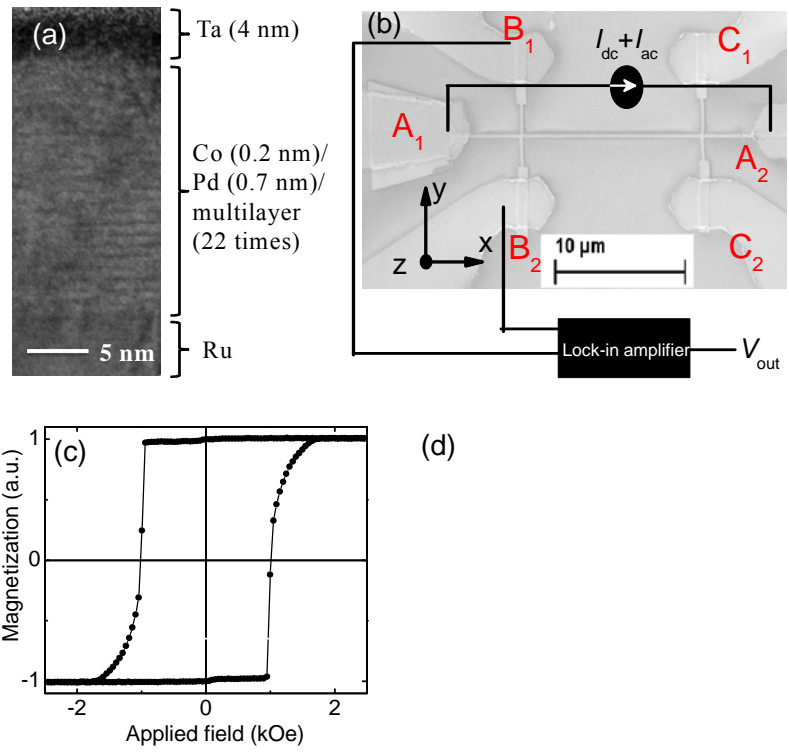

Figure 1

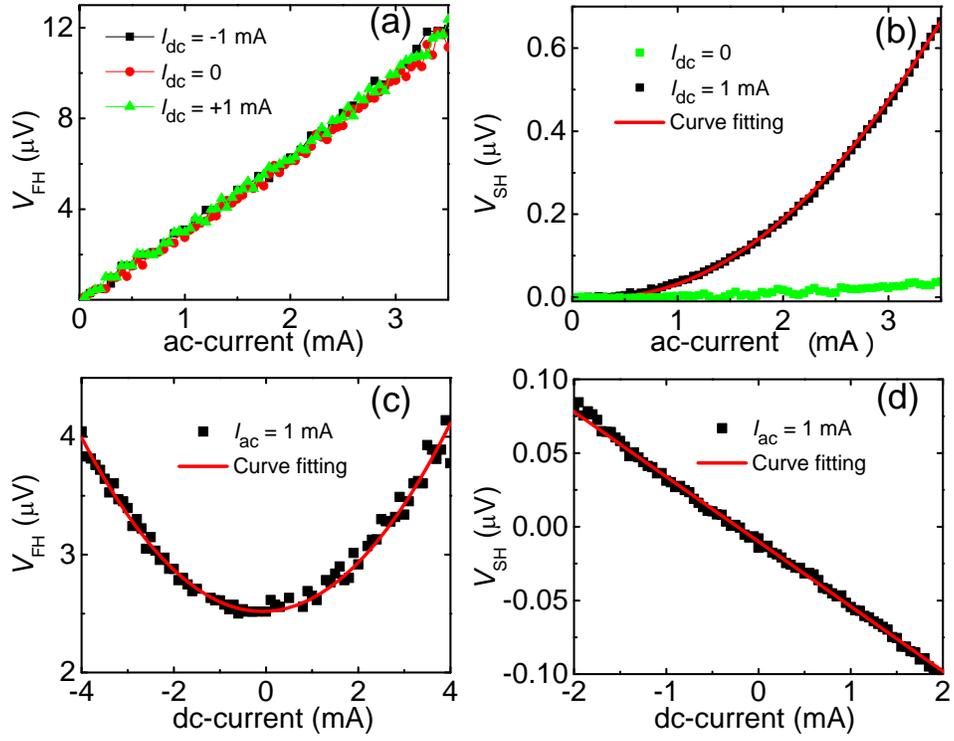

Figure 2

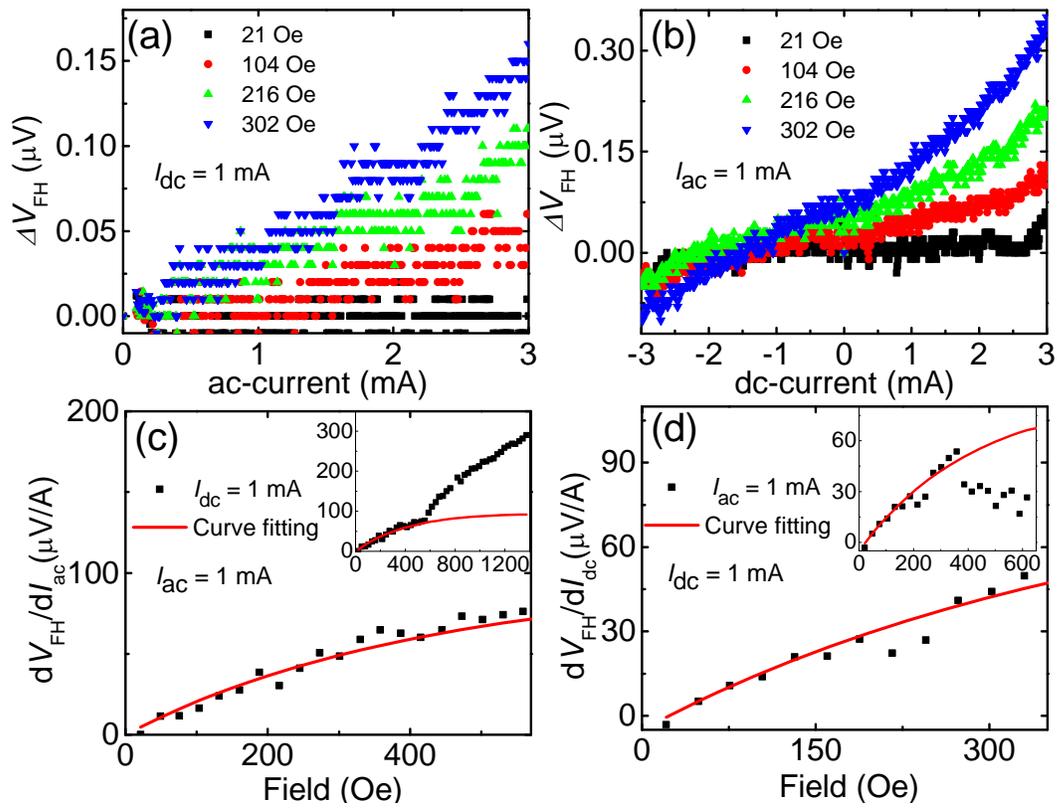

Figure 3

Table 1: Summary of the reported longitudinal and transverse torque components and the extracted dimensionless coefficients. The values in the brackets indicate the corresponding effective efficiency $\alpha_{\parallel,\perp}$. For the present work we used $t = 20$ nm and $M_S = 6.23\times10^5$ A/m. Note that the torques from Ref. 27 are taken at $\theta=0$.

| Structure (nm) | $\beta_\parallel$ (Oe/$10^8$ A/cm$^2$) [$\alpha_\parallel$] | $\beta_\perp$ (Oe/$10^8$ A/cm$^2$) [$\alpha_\perp$] | $\beta_\perp/\beta_\parallel$ | Ref. |
|---|---|---|---|---|
| Ta(4)/Co$_{40}$Fe$_{40}$B$_{20}$(1)/MgO(1.6) | 350 [0.12] | - | | [6] |
| Ta(3)/Co$_{40}$Fe$_{40}$B$_{20}$(0.9)/MgO(2) | 240 [0.07] | 450 [0.13] | 1.9 | [27] |
| Ta(1.5)/Co$_{40}$Fe$_{40}$B$_{20}$(1)/MgO(1.6) | 135 [0.078] | 472 [0.27] | 4 | [10] |
| Pt(3)/Co(0.6)/AlO$_x$(1.6) | 690 [0.13] | 400 [0.073] | 0.58 | [27] |
| Ta(4)/Ru(20)/(Co/Pd)$_{22}$/Ta(4) | 1170 [4.4] | 5025 [19.1] | 4.3 | This work |